\documentclass[prd,aps,preprint,amsmath,amssymb,eshowkeys,nofootinbib]{revtex4-1}
\usepackage{verbatim,graphics,graphicx,color,slashed,textcomp,bbm,bm}
\usepackage[normalem]{ulem} 
\usepackage[colorlinks=true,linkcolor=red,urlcolor=blue,citecolor=blue]{hyperref}

\newcommand{\GeV}{\mathrm{GeV}}

\begin{document}

\title{Weyl Scaling Invariant $R^2$ Gravity for Inflation and Dark Matter}
\author{Yong Tang$^{a,b,c}$ and Yue-Liang Wu$^{c,d,e,f}$}
\affiliation{\begin{footnotesize}
	${}^a$School of Astronomy and Space Sciences, University of Chinese Academy of Sciences (UCAS), Beijing, China\\
	${}^b$National Astronomical Observatories, Chinese Academy of Sciences, Beijing, China \\
	${}^c$School of Fundamental Physics and Mathematical Sciences, \\
	Hangzhou Institute for Advanced Study, UCAS, Hangzhou 310024, China \\
	${}^d$International Centre for Theoretical Physics Asia-Pacific, Beijing/Hangzhou, China \\
	${}^e$Institute of Theoretical Physics, Chinese Academy of Sciences, Beijing 100190, China \\
	${}^f$School of Physical Sciences, University of Chinese Academy of Sciences, Beijing, China   \end{footnotesize}}

\begin{abstract}
Inflation in the early universe can generate the nearly conformal invariant fluctuation that leads to the structures we observe at the present. The simple viable Starobinsky $R^2$ inflation has an approximate global scale symmetry. We study the conformal symmetric Weyl $\hat{R}^2$ and general $F(\hat{R})$ theories and demonstrate their equivalence to Einstein gravity coupled with a scalar and a Weyl gauge field. The scalar field in Weyl $\hat{R}^2$ gravity can be responsible for inflation with Starobinsky model as the attractor, potentially distinguishable from the latter by future experiments. The intrinsic Weyl gauge boson becomes massive once the Einstein frame is fixed, and constitutes as a dark matter candidate with mass up to $\sim 5\times 10^{16}\GeV$. 
\end{abstract}	

\maketitle

\section{Introduction}
The paradigm that there was an inflationary period in the early universe provides a compelling solution to the horizon and flatness problems~\cite{Starobinsky:1980te, Guth:1980zm, Linde:1981mu, Albrecht:1982wi}, and also explains the primordial origin for the almost scale-invariant density fluctuation that is responsible for the observed large scale structure in the late universe~\cite{Mukhanov:1981xt}. Although the exact model for inflation is not known yet, ongoing and  future planned experiments will give important information, for example, the spectral index and the strength of primordial gravitational waves, which would be useful to distinguish different models. 

The usual Einstein gravity assisted with a $R^2$ term stands out as a simple, elegant and well motivated inflation model, known as the Starobinsky inflation~\cite{Starobinsky:1980te} that was proposed originally to avoid the singularity problem. This model has an approximate global scale invariance because the Einstein-Hilbert term $R$ explicitly breaks the scaling symmetry, and is equivalent to adding a new scalar~\cite{Stelle:1977ry, Whitt:1984pd}. Extending the global scaling symmetry into local Weyl conformal symmetry results in new interactions with additional physical degrees of freedom, as shown firstly by Weyl in~\cite{Weyl:1929fm} where a new gauge field was introduced. Since proposed, Weyl conformal symmetry with various studies have been explored in gauge theory of quantum gravity~\cite{Wu:2015wwa, Wu:2017urh}, induced gravity~\cite{Zee:1978wi, Adler:1982ri, Fujii:1982ms, Salvio:2014soa, Oda:2020yyv}, scale-invariant extensions of the standard model of particle physics~\cite{Cheng:1988zx, Hur:2011sv, Holthausen:2013ota, Foot:2007iy, Nishino:2009in, Farzinnia:2013pga, Guo:2015lxa, Kubo:2018kho}, inflation and late cosmology~\cite{Wetterich:1987fm, Kaiser:1994vs, Wu:2004rs, Ferrara:2010in, GarciaBellido:2011de,  Giudice:2010ka, Kurkov:2013gma, Bars:2013yba, Csaki:2014bua, Kannike:2015apa, Kannike:2015kda, Salvio:2017xul, Ferreira:2018qss, Tang:2018mhn, Tang:2019uex, Tang:2019olx, Barnaveli:2018dxo, Ghilencea:2018thl, Ghilencea:2018dqd, Ghilencea:2019rqj,  Ferreira:2019zzx, Gunji:2019wtk, Ishiwata:2018dxg, Ishida:2019wkd, Barnaveli:2018dxo, Gialamas:2019nly, Karam:2018mft, Ghilencea:2020piz}. 

In this paper we study inflation and dark matter (DM) in the Weyl symmetric $\hat{R}^2$ and general $F(\hat{R})$ gravity, where $\hat{R}$ is the modified Ricci scalar that contains the Weyl gauge field intrinsically. We show these theories are equivalent to Einstein gravity coupled with a new scalar degree of freedom that can be responsible for inflation, and a Weyl gauge boson that can be a dark matter candidate. We also find that generally the signs of the kinetic term of scalars in these theories can be positive or negative, both of which allow analytic treatments. More intriguingly, we demonstrate that the inflationary observables in $\hat{R}^2$ are different from Starobinsky model, but with later as an attractor~\footnote{This point was firstly discussed in~\cite{Ghilencea:2019rqj, Ferreira:2019zzx}. We provide an analytical formalism, see texts for details.}, which can be tested in future experiments with sensitivity of tensor-to-scalar ratio $r\sim 0.003$. Once the Einstein frame is fixed, the intrinsically associated Weyl gauge boson becomes massive as it absorbs one scalar as its longitudinal model, and is a possible DM candidate~\footnote{DM in this model is different from other scenarios where inflaton is identified as DM~\cite{Kofman:1997yn, Lerner:2009xg, Mukaida:2013xxa, Khoze:2013uia, Hooper:2018buz, Borah:2018rca, Choi:2019osi, Daido:2017tbr, Daido:2017wwb, Tenkanen:2016twd}, and is also unlike~\cite{Katsuragawa:2016yir} where DM is identified as the scalaron in $f(R)$ gravity.}, with mass up to $\sim 5\times 10^{16}\GeV$. 

This paper is organized as follows. In Sec.~\ref{sec:general}, we establish the conventions and general formalism for Weyl gravity. Then in Sec.~\ref{sec:Weyl}, we discuss the details of Weyl $\hat{R}^2$ gravity and illustrate how it connects with Einstein gravity that couples to a scalar field and a Weyl gauge field. We further extend the formalism to theories with functions of $\hat{R}$ and multiple scalars. Later in Sec.~\ref{sec:inflation} we demonstrate how the scalar can act as the inflaton field and compare it with the Starobinsky model. In Sec.~\ref{sec:dark} we briefly discuss the physics of Weyl gauge boson as dark matter in this scenario. Finally, we give our conclusions. 

Throughout the paper, we use the metric with a sign convention $(-1,+1,+1,+1)$, and the natural unit, $\hbar=c=1, M_{p}\equiv 1/\sqrt{ 8\pi G}=1$. Greek letters, $\mu,\nu,...$, refer to spacetime indices, $(0,1,2,3)$. 

\section{Notation and Formalism}\label{sec:general}
Before discussing the Weyl $\hat{R}^2$ gravity, let us first consider the following action for metric tensor field $g_{\mu\nu}$, Weyl gauge field $W_\mu\equiv g_W w_\mu$ and scalar field $\phi$, 
\begin{equation}\label{eq:lag}
	S=\int d^4 x \mathcal{L},\ \mathcal{L}= \sqrt{-g}\left[\frac{1}{2}\phi^{2}\hat{R}-\frac{1}{2} \zeta D^\mu \phi D_{\mu}\phi-\frac{1}{4g^2_W}F_{\mu\nu}F^{\mu\nu}-\lambda \phi^4\right], 
\end{equation}
where $g_W$ is the gauge coupling associated with Weyl gauge symmetry defined shortly below, the covariant derivative on scalar $D_\mu = \partial_{\mu}-W_\mu$, and the modified Ricci scalar $\hat{R}$ is obtained from $\hat{R}_{\;\sigma\mu\nu}^{\rho}$, 
\begin{eqnarray}
\hat{R}_{\;\sigma\mu\nu}^{\rho}&=&\partial_{\mu}\hat{\Gamma}_{\sigma\nu}^{\rho}-\partial_{\nu}\hat{\Gamma}_{\sigma\mu}^{\rho}+\hat{\Gamma}_{\mu\tau}^{\rho}\hat{\Gamma}_{\sigma\nu}^{\tau}-\hat{\Gamma}_{\nu\tau}^{\rho}\hat{\Gamma}_{\sigma\mu}^{\tau},\
\hat{R}_{\sigma\nu}=\hat{R}_{\;\sigma\rho\nu}^{\rho},\ \hat{R}=g^{\sigma\nu}\hat{R}_{\sigma\nu}, \\
\hat{\Gamma}_{\mu\nu}^{\rho}&=&\Gamma_{\mu\nu}^{\rho}+\left[W_{\mu}\delta_{\nu}^{\rho}+W_{\nu}\delta_{\mu}^{\rho}-W^{\rho}g_{\mu\nu}\right],
\Gamma_{\mu\nu}^{\rho}=\frac{1}{2}g^{\rho\sigma}\left(\partial_{\mu}g_{\sigma \nu} + \partial_{\nu}g_{\mu \sigma} - \partial_{\sigma}g_{\mu\nu}\right). 
\end{eqnarray}
Note that $\hat{\Gamma}_{\mu\nu}^{\rho}$ can be derived when we replace the usual derivatives on metric tensor in Christoffel symbol $\Gamma_{\mu\nu}^{\rho}$ with 
\begin{equation}
\partial_{\mu}g_{\rho\sigma}\rightarrow\left(\partial_{\mu}+2W_{\mu}\right)g_{\rho\sigma}.
\end{equation}
We have kept a parameter $\zeta$ in the front of kinetic term of $\phi$, which can be positive, negative and zero. The point that $\zeta$ can be negative was largely overlooked in some previous studies~\cite{Ghilencea:2019rqj, Ferreira:2019zzx, Oda:2020yyv} and we have showed consistent theories are possible with both signs in \cite{Tang:2019uex, Tang:2019olx}. The reason for this introduction will be clear in later discussions. Since we have the freedom to rescale $\phi$, we can normalize the coefficient in the front of $\phi^2\hat{R}$ to be $1/2$. Therefore the absolute value $|\zeta|$ is not theoretically bounded at the point. Note that the theory with positive $\zeta$ can not be transformed to one with negative $\zeta$, by redefinition of $\phi$, especially in the context where $\phi$ is also coupled to other scalar and fermion fields, as we shall show later.

Under the following Weyl conformal transformation parametrized by a positive transfer function $f(x)$,
\begin{eqnarray}\label{eq:weyl}
g_{\mu\nu}&\rightarrow&{g}'_{\mu\nu}=f^{2}g_{\mu\nu}, \\
W_{\mu}&\rightarrow&{W}'_{\mu}=W_{\mu}-\partial_{\mu}\ln f,\\
\phi &\rightarrow&{\phi}'=f^{-1}\phi,
\end{eqnarray}
the Lagrangian $\mathcal{L}$ is invariant, namely, independent on $f$. More precisely speaking, each term in the bracket of eq.~(\ref{eq:lag}) ($\times\sqrt{-g}$) is invariant. This can be easily verified by using the following identities,
\begin{align}
\hat{\Gamma}_{\mu\nu}^{'\rho}&=\hat{\Gamma}_{\mu\nu}^{\rho},\;\hat{R}_{\;\sigma\mu\nu}^{'\rho}=\hat{R}_{\;\sigma\mu\nu}^{\rho},\ D'_\mu \phi' =fD_\mu \phi . 
\end{align}

Although tedious, it is straightforward to show the relation between $\hat{R}$ and $R$, 
\begin{equation}\label{eq:rterm}
\hat{R}=R-6W_{\mu}W^{\mu}-6\nabla_{\mu}W^{\mu},\ \nabla_{\mu}W^\mu = \frac{6}{\sqrt{-g}}\partial_{\mu}\left(\sqrt{-g}W^{\mu}\right),
\end{equation}
where $R$ is usual Ricci scalar defined by ${\Gamma}_{\mu\nu}^{\rho}$. With this quantity, we can rewrite 
\begin{align}
\phi^{2}\hat{R} & = \phi^{2}R-6\phi^{2}W_{\mu}W^{\mu}-\frac{6\phi^{2}}{\sqrt{-g}}\partial_{\mu}\left(\sqrt{-g}W^{\mu}\right),\nonumber \\
&=\phi^{2}R-6\phi^{2}W_{\mu}W^{\mu}+12\phi\partial_{\mu}\phi W^{\mu}-\frac{6}{\sqrt{-g}}\partial_{\mu}\left(\sqrt{-g}\phi^{2}W^{\mu}\right),\nonumber \\
&\rightarrow \phi^{2}R-6\left(\partial_{\mu}\phi-W_{\mu}\phi\right)^{2}+6\partial_{\mu}\phi\partial^{\mu}\phi= \phi^{2}R+6\partial_{\mu}\phi\partial^{\mu}\phi-6D_{\mu}\phi D^{\mu}\phi.
\end{align}
In the last line above, we have dropped the surface term which vanishes in our interested cases. The combination $\phi^{2}R+6\partial_{\mu}\phi\partial^{\mu}\phi$ is conformal invariant and widely used in the literature. Note that $D_{\mu}\phi D^{\mu}\phi$ can be joined with $\zeta$ term with $\zeta\rightarrow \zeta + 6$, which is one of the reasons why we have kept $\zeta$ apparent. 

Since the theory is conformal invariant, it has the freedom to fix the frame or gauge by choosing proper $f$ in the Weyl transformation. To compare with the Einstein gravity, it is physically natural to adopt $f(x)=\phi$ in the Weyl transformation eq.~(\ref{eq:weyl}) or equivalently fix $\phi^2=1$, then we would restore Einstein gravity~\footnote{Throughout our paper, all physical observables are defined in the final Einstein frame, where dimensional quantities can arise due to the breaking of scaling invariance. In general scalar-tensor theories without scaling symmetry, Einstein frame and Jordan frame is physically equivalent at linear order~\cite{Fujii:2003pa}.} with a massive Weyl gauge boson $W_\mu$,
\begin{equation}
\mathcal{L}= \sqrt{-g}\left[\frac{1}{2}R-\lambda-\frac{1}{4g^2_W}F_{\mu\nu}F^{\mu\nu}-\frac{1}{2}\left(\zeta+6\right)W_\mu W^\mu\right],
\end{equation}
where the mass for $W_\mu$ is given by $g_W\sqrt{\zeta + 6}$ and $\zeta+6>0$ is required to have positive and real mass for $W_\mu$. It is seen that the physical degree of freedom $\phi$ has been absorbed as the longitudinal model of $W_\mu$. Because there is a $Z_2$ symmetry for $W_\mu$, $W_\mu \rightarrow - W_\mu$, Weyl gauge boson $W_\mu$ is stable and can be a good dark matter candidate when having the correct relic abundance. We have discussed this scenario extensively in Ref.~\cite{Tang:2019uex,Tang:2019olx}, where we have also proved that the conclusion is also true for theories with multiple scalars. 

There is a cosmological constant $\lambda$ in above formalism, whose value can be determined only by experimental or observational data. If it is identified as the current dark energy density, $\lambda$ should be as small as $10^{-120}M^4_P$. In such a case, $\lambda$ term is only important in the very late universe when it dominates the energy density and can be omitted in discussing physical effects in the early inflationary cosmology. In principle, the effect of this $\lambda$ term is different in other model setup, which we shall show in Weyl $\hat{R}^2$ gravity in the next section.

\section{Weyl $\hat{R}^2$ Gravity}\label{sec:Weyl}
\subsection{$\hat{R}^2$ Gravity}
In this section, we discuss the Lagrangian with an additional $\hat{R}^2$ term,
\begin{equation}\label{eq:R2}
\mathcal{L}= \sqrt{-g}\left[\frac{1}{2}\phi^{2}\hat{R}+\frac{\alpha}{12}\hat{R}^2-\frac{1}{2} \zeta D^\mu \phi D_{\mu}\phi-\frac{1}{4g^2_W}F_{\mu\nu}F^{\mu\nu}-\lambda \phi^4\right]. 
\end{equation}
Inflation physics in this model is also discussed in Refs.~\cite{Ferreira:2019zzx, Ghilencea:2019rqj}. Here, we have obtained several important new insights: a) Even if $\lambda=0$, inflation is possible and viable. b) theories with negative $\zeta$ are essentially different from these with positive $\zeta$, and both are consistent and can not be transformed to each other by redefinition of $\phi$. c) Our treatment is also applied for theories with general $F(\hat{R})$ and Weyl symmetry. d) Analytic demonstration of Starobinsky inflation as an attractor is provided. e) The Weyl gauge boson can be a dark matter candidate, with mass up to $\sim 5\times 10^{16}\GeV$, through scalar annihilation after inflation. 

There are several reasons why we only add the $R^2$ term here. Firstly, the usual $R^2$ term can give a viable inflation model, namely, the Starobinsky inflation~\cite{Starobinsky:1980te}. This makes it an attractive candidate for extension of Einstein's gravity and it would be useful to compare with this benchmark model. Secondly, it is the lowest-order conformally invariant term that can be added without unitarity and instability issues. The introduction of $\hat{R}_{\mu\nu}\hat{R}^{\mu\nu}$ and  $\hat{R}_{\mu\nu\rho\sigma}\hat{R}^{\mu\nu\rho\sigma}$ generally would be accompanied with instabilities caused by other new degrees of freedom~\cite{Stelle:1976gc, Stelle:1977ry, Woodard:2015zca}.

From eq.~(\ref{eq:rterm}), at first sight we can observe that $\hat{R}^2$ would introduce more terms involving $W_\mu$, such as cubic and quartic ones,
\begin{equation}
\hat{R}^2=R^{2}+36\left(\nabla_{\mu}W^{\mu}+W_{\mu}W^{\mu}\right)^{2}-12R\left(\nabla_{\mu}W^{\mu}+W_{\mu}W^{\mu}\right).
\end{equation}
Then, one may naively expect that the $Z_2$ symmetry mentioned in the last section would not be preserved any more. Furthermore, since a new kinetic term $\nabla_{\mu}W^{\mu}\nabla_{\nu}W^{\nu}$ appears differently from the usual massive vector theories, it seems additional new vectorial degree of freedom are introduced. In the following, we shall how both naive expectations are not the real case. Note that we may add to the Lagrangian a particular combination,
\begin{equation}
\frac{1}{2}W_{\mu\nu\rho\sigma}W^{\mu\nu\rho\sigma}=\frac{1}{6}\hat{R}^2-\hat{R}_{\mu\nu}\hat{R}^{\mu\nu}+\frac{1}{2}\hat{R}_{\mu\nu\rho\sigma}\hat{R}^{\mu\nu\rho\sigma}=\frac{1}{6}{R}^2-{R}_{\mu\nu}{R}^{\mu\nu}+\frac{1}{2}{R}_{\mu\nu\rho\sigma}{R}^{\mu\nu\rho\sigma},
\end{equation}
which is independent of $W_\mu$. Here Weyl tensor is defined by
\begin{equation}
W_{\mu\nu\rho\sigma}=\hat{R}_{\mu\nu\rho\sigma}-g_{\mu[\rho}\hat{R}_{\sigma]\nu}+g_{\nu[\rho}\hat{R}_{\sigma]\mu}+\frac{1}{3}\hat{R}g_{\mu[\rho}g_{\sigma]\nu}.
\end{equation} 
However, since high-derivative terms from $\hat{R}_{\mu\nu}\hat{R}^{\mu\nu}$ and  $\hat{R}_{\mu\nu\rho\sigma}\hat{R}^{\mu\nu\rho\sigma}$ still appear, the theory would again suffer unitarity and instability issues. Therefore, we only focus on $\hat{R}^2$ for the rest of our discussions.

We introduce an auxiliary field $\chi$, and rewrite the Lagrangian eq.~(\ref{eq:R2}) as 
\begin{equation}\label{eq:linear}
\mathcal{L}= \sqrt{-g}\left[\frac{1}{2}\left(\phi^2+\frac{\alpha}{3}\chi^2\right)\hat{R}-\frac{\alpha}{12}\chi^4-\frac{1}{2} \zeta D^\mu \phi D_{\mu}\phi-\frac{1}{4g^2_W}F_{\mu\nu}F^{\mu\nu}-\lambda \phi^4\right]. 
\end{equation}
The equivalence can be shown after we use the equation of motion for $\chi$, 
\begin{equation}
\frac{\delta \mathcal{L}}{\delta \chi}=0\Rightarrow \chi^2=\hat{R},
\end{equation}
and put it back into eq.~(\ref{eq:linear}). This also explicitly demonstrates that introduction of $\hat{R}^2$ is equivalent to a new scalar degree of freedom. 

Note that the Lagrangian still respects the Weyl symmetry, therefore we have the freedom to choose the transfer function $f$ to fix the frame. In this case, we shall use $f^2=\phi^2+\frac{\alpha}{3}\chi^2$ or equivalently set $\phi^2+\frac{\alpha}{3}\chi^2=1$ to restore the Einstein gravity. Then, we get
\begin{equation}\label{eq:after}
\frac{\mathcal{L}}{\sqrt{-g}}=\frac{1}{2}R-\frac{1}{2} \zeta D^\mu \phi D_{\mu}\phi-\frac{1}{4g^2_W}F_{\mu\nu}F^{\mu\nu}-3W^\mu W_\mu-\frac{3}{4\alpha}\left(1-\phi^2\right)^2-\lambda \phi^4. 
\end{equation}
Noting that there is a linear term of $W_\mu$ in the kinetic term of $\phi$, we can rewrite 
\begin{align}\label{eq:gauge}
\frac{1}{2}\zeta D^\mu \phi D_{\mu}\phi + 3W^\mu W_\mu =&\frac{1}{2}\left[\left(6+\zeta \phi^2\right)W^\mu W_\mu-\zeta W^\mu \partial_\mu \phi^2\right]+\frac{1}{2}\zeta \partial^\mu \phi \partial_{\mu}\phi \nonumber \\
=& \frac{1}{2}\left(6+\zeta \phi^2\right)\left[W_\mu - \frac{1}{2}\partial_\mu \ln \left(6+\zeta \phi^2\right)\right]^2 + \frac{1}{2}\frac{6\zeta}{6+\zeta \phi^2}\partial^\mu \phi \partial_{\mu}\phi,
\end{align}
and make the redefinition
\begin{equation}
\overline{W}_\mu = W_\mu - \frac{1}{2}\partial_\mu \ln \left(6+\zeta \phi^2\right),
\end{equation}
which is nothing but an usual gauge transformation for $W_\mu$ and would not change the kinetic term, $F_{\mu\nu}F^{\mu\nu}=\overline{F}_{\mu\nu}\overline{F}^{\mu\nu}$. As a result, we have the following Lagrangian, 
\begin{equation}\label{eq:inflag}
\frac{\mathcal{L}}{\sqrt{-g}}= \frac{1}{2}R-\frac{1}{4g^2_W}\overline{F}_{\mu\nu}\overline{F}^{\mu\nu}-\frac{1}{2}\left(6+\zeta \phi^2\right)\overline{W}^\mu \overline{W}_\mu-\frac{3\zeta}{6+\zeta \phi^2}\partial^\mu \phi \partial_{\mu}\phi-\frac{3}{4\alpha}\left(1-\phi^2\right)^2-\lambda \phi^4. 
\end{equation}
Now it becomes clear that $Z_2$ symmetry is restored for $\overline{W}_\mu$. 

As long as the relation $\zeta\left(6+\zeta \phi^2\right)>0$ is satisfied, we would have a normal scalar. Then we can define a new scalar field $\sigma$ by
\begin{equation}
	\frac{d\sigma }{d\phi }= \pm \sqrt{\frac{6\zeta }{6+\zeta \phi^2}}.
\end{equation}
Then $\sigma(x)$ would have the canonical kinetic terms. For the $\zeta=0$, $\sigma$ or $\phi$ is not dynamical and its value can be fixed by the equation of motion, which is a trivial case that we shall not discuss any further. For $\zeta\neq 0$, we have the analytic solutions,
\begin{align}
\phi &= \sqrt{\frac{6}{+\zeta}}\sinh \frac{\pm \sigma }{\sqrt{6}}, \;\textrm{ for }\zeta>0,\label{eq:sol1}\\
\phi &= \sqrt{\frac{6}{-\zeta}}\cosh \frac{\pm \sigma }{\sqrt{6}}, \;\textrm{ for }\zeta<0.\label{eq:sol2}
\end{align}
Finally, we have a theory of Einstein gravity, a canonical scalar $\sigma$ and a massive vector $\overline{W}_\mu$,
\begin{equation}\label{eq:ein}
\frac{\mathcal{L}}{\sqrt{-g}}= \frac{1}{2}R-\frac{1}{4g^2_W}\overline{F}_{\mu\nu}\overline{F}^{\mu\nu}-\frac{1}{2}\left(6+\zeta \phi^2\right)\overline{W}^\mu \overline{W}_\mu-\frac{1}{2}\partial^\mu \sigma \partial_{\mu}\sigma-\frac{3}{4\alpha}\left(1-\phi^2\right)^2-\lambda \phi^4,
\end{equation}
where $\phi$ now is a function of $\sigma$ in eqs.~(\ref{eq:sol1}) and (\ref{eq:sol2}), depending on the sign of $\zeta$. The scalar potential is minimized at $\phi^2 = 1/(1+4\lambda\alpha/3)$ with $V=0$. Note that the special case with $\zeta=6$ and positive sign in the solution was studied in~\cite{Ghilencea:2019rqj}, whose results we agree with. 

In the limit with $\zeta\rightarrow \infty$, we have the approximation $6+\zeta\phi^2\simeq \zeta\phi^2$ for non-zero $\phi$ in eq.~\ref{eq:inflag}, then we would have
\begin{equation}
 \frac{d\sigma}{d\phi} = \pm \frac{\sqrt{6}}{\phi},
\end{equation}
whose solution is $\phi = \exp(\pm \sigma/\sqrt{6})$ and independent on $\zeta$. Put the solution with negative sign into the potential $(1-\phi^2)^2$, we get exactly the same potential in Starobinsky model, which is given by
\begin{equation}\label{eq:starobinsky}
\mathcal{V}\left(\sigma \right)=\frac{3}{4\alpha} \left[1-\exp\left(-\sqrt{\frac{2}{3}}\sigma\right)\right]^{2}.
\end{equation}
In the large $\zeta$ limit, $\overline{W}_\mu$ has an infinitely large mass and will decouple from other fields. Then, this theory effectively reduces to Starobinsky model. For inflationary observables, the equivalence depends on whether $\zeta \phi^2 \gg 6$ is satisfied or not in the relevant field interval during inflation, about which we shall discuss more in section~\ref{sec:inflation}. 

Note that the attractor behavior to Starobinsky model as $\zeta\rightarrow \infty$ would not appear if we take $\zeta\rightarrow \infty$ from the very beginning, namely, from eq.~(\ref{eq:after}), in which case both $\phi$ and $W_\mu$ is decoupled. The usual lore is that when a term in the Lagrangian is taking to be infinity large, it would decouple from other fields and therefore can be neglected. However, this wisdom does not apply here because there is a mixing term between $W_\mu$ and $\phi$. After a gauge transformation of $W_\mu$, a new contribution of kinetic term arises for $\phi$ and a finite sum as $\zeta\rightarrow \infty$ is left over. This is clearly illustrated in eq.~(\ref{eq:gauge}). The lesson learned here is that we should always work with the final Lagrangian that has $Z_2$ symmetry for the physical $\overline{W}_\mu$. 

\subsection{$F(\hat{R},\phi)$ Gravity }
The above formalism can be extended to models with functions of $\hat{R}$ and $\phi$. If the Lagrangian has the following form
\begin{equation}
\frac{\mathcal{L}}{\sqrt{-g}}= \frac{1}{2}F(\hat{R},\phi)-\frac{1}{2} \zeta D^\mu \phi D_{\mu}\phi-\frac{1}{4g^2_W}F_{\mu\nu}F^{\mu\nu}-\lambda \phi^4. 
\end{equation}
where $F(\hat{R},\phi)$ is a function of $\hat{R}$ without derivatives and $\phi$ with dimensional four to be conformal invariant. In our discussion above, $F(\hat{R},\phi)=\phi^2 \hat{R} + \alpha \hat{R}^2/6$. We can similarly use the auxiliary field $\chi$ and rewrite the Lagrangian
\begin{equation}\label{eq:fr}
\frac{\mathcal{L}}{\sqrt{-g}}= \frac{1}{2}\left[F(\chi^2,\phi)+F_{\hat{R}}\left(\chi^2,\phi\right)\left(\hat{R}-\chi^2\right)\right]-\frac{1}{2} \zeta D^\mu \phi D_{\mu}\phi-\frac{1}{4g^2_W}F_{\mu\nu}F^{\mu\nu}-\lambda \phi^4,
\end{equation}
where $F_{\hat{R}}$ denotes the derivative of $F(\hat{R},\phi)$ over $\hat{R}$ and $F(\chi^2,\phi)\equiv F(\hat{R}\rightarrow \chi^2,\phi)$. 
One may check that the equation of motion for $\chi$ still gives $\chi^2=\hat{R}$. 

It is also apparent that the above new Lagrangian is conformal invariant and linear on $\hat{R}$. Likewise, we can choose the conformal transformation with $f\equiv F_{\hat{R}} \left(\chi^2,\phi \right)=1$ and the resulting theory describes Einstein gravity coupled with a scalar and a massive vector and has the same Lagrangian as eq.~(\ref{eq:ein}) but with the potential replaced by $\mathbf{V}(\phi)=-\left[F(\chi^2,\phi)-\chi^2 F_{\hat{R}}\left(\chi^2,\phi\right)\right]/2+\lambda \phi^4$. Once again, $Z_2$ symmetry for the Weyl gauge boson is preserved.

Although the formalism is straightforward for any $F(\hat{R},\phi)$, in reality the scalar potential is very complicated even if we modify our previous case slightly. For instance, we may consider the following function with $\hat{R}^3$,
\begin{equation}
F(\hat{R},\phi)=\phi^2 \hat{R} + \frac{\alpha}{6}\hat{R}^2 + \frac{\beta}{6\phi^2}\hat{R}^3.
\end{equation}
Then we can work out 
\begin{align}
	F_{\hat{R}} \left(\chi^2,\phi\right) &= \phi^2 +  \frac{\alpha}{3}\chi^2 + \frac{\beta}{2\phi^2}\chi^4,\\
	F\left(\chi^2,\phi\right)&=\phi^2 \chi^2 + \frac{\alpha}{6}\chi^4 + \frac{\beta}{6\phi^2}\chi^6. 
\end{align}
The scalar potential is given by 
\begin{equation}
\mathbf{V}(\phi)=\frac{\alpha}{6}\chi^4+\frac{\beta}{3\phi^2}\chi^6+\lambda \phi^4.
\end{equation}
where $\chi^2$ as a function of $\phi$ is determined by $F_{\hat{R}} \left(\chi^2,\phi \right)=1$, $ \frac{\beta}{2}\chi^4+\frac{\alpha}{3}\phi^2\chi^2 -\phi^2(1-\phi^2)=0$,
\begin{equation}
\chi^2 = \frac{	\alpha/3 \pm \sqrt{\alpha^2/9+2\beta (1-\phi^2)/\phi^2}}{\beta }\phi^2.
\end{equation}
The resulting $\mathbf{V}(\phi)$ is a complicated function of $\phi$ which itself depends on the canonical scalar $\sigma$ as in previous section. In principle, by choosing proper parameters, $\alpha, \beta$ and $\lambda$, inflation region shall exist in this theory, for instance, $\beta\rightarrow 0$. However, there is no transparently new insights in this case. Therefore, we shall restrict to $\hat{R}^2$ case in our later discussions about inflation and dark matter physics. 

\subsection{Multiple Scalars}
We have so far only discussed the case with a single scalar $\phi$. One may wonder what happens if we have $N$ scalars that couple to gravity non-minimally. For example, the Lagrangian can be given as following,
\begin{equation}
\frac{\mathcal{L}}{\sqrt{-g}}=\frac{1}{2}\hat{R}\sum_{I}\beta_{I}\phi_{I}^{2}+\frac{\alpha}{12}\hat{R}^{2}-\frac{1}{2}\sum_{I}\zeta_{I}D^{\mu}\phi_{I}D_{\mu}\phi_{I}-\frac{1}{4g_{W}^{2}}F_{\mu\nu}F^{\mu\nu}-V\left(\phi_{I}\right).
\end{equation}
Here we have kept $\beta_{I}$ and now it may have three values, $\beta_I=\pm 1$ or $0$, $I=1,2,...,N$. The Weyl invariant scalar potential has a form
\begin{equation}
V\left(\phi_I\right)=\sum_{i,j,...,n}\sum_{I,J,...,N}C[{\phi_K}/{\phi_L}]\phi^i_{I}\phi^j_{J}\cdots \phi^n_{N}, \;i+j+...+n=4,
\end{equation}
where $C[{\phi_K}/{\phi_L}]$ are dimensionless real functions of the ratios ${\phi_K}/{\phi_L}$. It is clear that $\phi_I\rightarrow i \phi_I$ could change the sign of $\zeta_{I}$ but make the potential and Lagrangian imaginary, an unacceptable situation. 

Using similar procedures, we can obtain the Lagrangian in Einstein frame, 
\begin{align}
\frac{\mathcal{L}}{\sqrt{-g}}=&
\frac{1}{2}R-\frac{1}{4g_{W}^{2}}\overline{F}_{\mu\nu}\overline{F}^{\mu\nu}-\frac{\left(6+\sum_{I}\zeta_{I}\phi_{I}^{2}\right)}{2}\overline{W}_{\mu}\overline{W}^{\mu}\nonumber -V\left(\phi_{I}\right)\\
&-\frac{1}{2}\frac{1}{6+\sum_{I}\zeta_{I}\phi_{I}^{2}}\sum_{I}\zeta_{I}\left[\left(6+\sum_{J\neq I}\zeta_{J}\phi_{J}^2\right)\partial_{\mu}\phi_{I}\partial^{\mu}\phi_{I}
-\sum_{J\neq I}\zeta_{J}\phi_{I}\phi_{J}\partial_{\mu}\phi_{I}\partial^{\mu}\phi_{J}\right]
.
\end{align}
In this case, $Z_2$ symmetry for $\overline{W}_\mu$ is also manifest. It can be easily checked that when $N=1$, it agrees with our previous result above. In such a general setup, non-canonical and mixed kinetic terms for $\phi_I$ are encountered. However, unlike the single scalar case, for $N>1$ generically we are unable to find field redefinition to get the canonically normalized kinetic terms. This may be easily understood from the following argument. For $N>1$, one can regard the metric for the field space with coordinates $\phi_I$, $G_{IJ}\left(\phi\right)$,
\begin{align}
G_{II}&=\zeta_{I} - \frac{\zeta^2_{I}\phi_{I}^2}{6+\sum_{J}\zeta_{J}\phi_{J}^{2}},\\
G_{IJ}&=-\frac{\zeta_{I}\zeta_{J}\phi_{I}\phi_{J}}{6+\sum_{I}\zeta_{I}\phi_{I}^{2}},\;\textrm{ for }I\neq J,
\end{align}
and the invariant length $ds^2=G_{IJ}d\phi_I d\phi_J$. If the corresponding Riemann tensor $\mathcal{R}^I_{\;JMN}=0$, we would have a flat geometry in the field space. Then there should exist field transformations, $\varphi_I=\varphi_I(\phi_{J})$, to have the new metric $\tilde{G}_{IJ}(\varphi)=\delta_{IJ}$ and $ds^2=\delta_{IJ}d\varphi_Id\varphi_J $. Since $\mathcal{R}^I_{\;JMN}\neq 0$ in our case for $N>1$, such transformations would be unlikely. Hence, a complete analytic treatment of the theory would be unrealistic for $N>1$. For the rest of this paper, we shall concentrate on $N=1$ case only.

\section{Inflation}\label{sec:inflation}
We are now in a position to discuss inflation. We shall demonstrate it is possible to have viable inflation in both cases, $\zeta>0$ and $\zeta<0$. When $\zeta<0$, we can use the relation, $\cosh^2 x - \sinh^2 x =1$, and rescale $\zeta$ and $\alpha$ to bring the potential into the same form as in $\zeta>0$. This suggests that as far as only inflation is concerned, these two cases give similar dynamics, although they are not equivalent to each other. Therefore, for inflation we only need to investigate the $\zeta>0$ case. For concrete discussions, we shall use the positive sign in the solution, eq.~(\ref{eq:sol1}). The potential from eq.~(\ref{eq:ein}) at inflation era is
\begin{equation}
V\left(\sigma \right)\simeq \frac{3}{4\alpha }\left[1-\frac{6}{\zeta}\sinh^2 \left(\frac{\sigma}{\sqrt{6}} \right)\right]^2.
\end{equation}
Here we have ignored the $\lambda$ term. When $\lambda$ is as small as we discussed in the end of sec.~(\ref{sec:general}), its contribution is negligible in the inflation epoch. However, it can still shift potential minimum to $\phi^2 = 1/(1+4\lambda\alpha/3)$ at which $V=0$, unlike the cosmological constant in sec.~(\ref{sec:general}). Another reason for the choice of small $\lambda$ is to compare transparently with the usual Starobinsky inflation. As we mentioned before, our scenario flows effectively into Starobinsky model $\zeta\rightarrow \infty$, which will be explicitly verified numerically below.

\begin{figure}[t]
	\includegraphics[width=0.49\textwidth,height=0.5\textwidth]{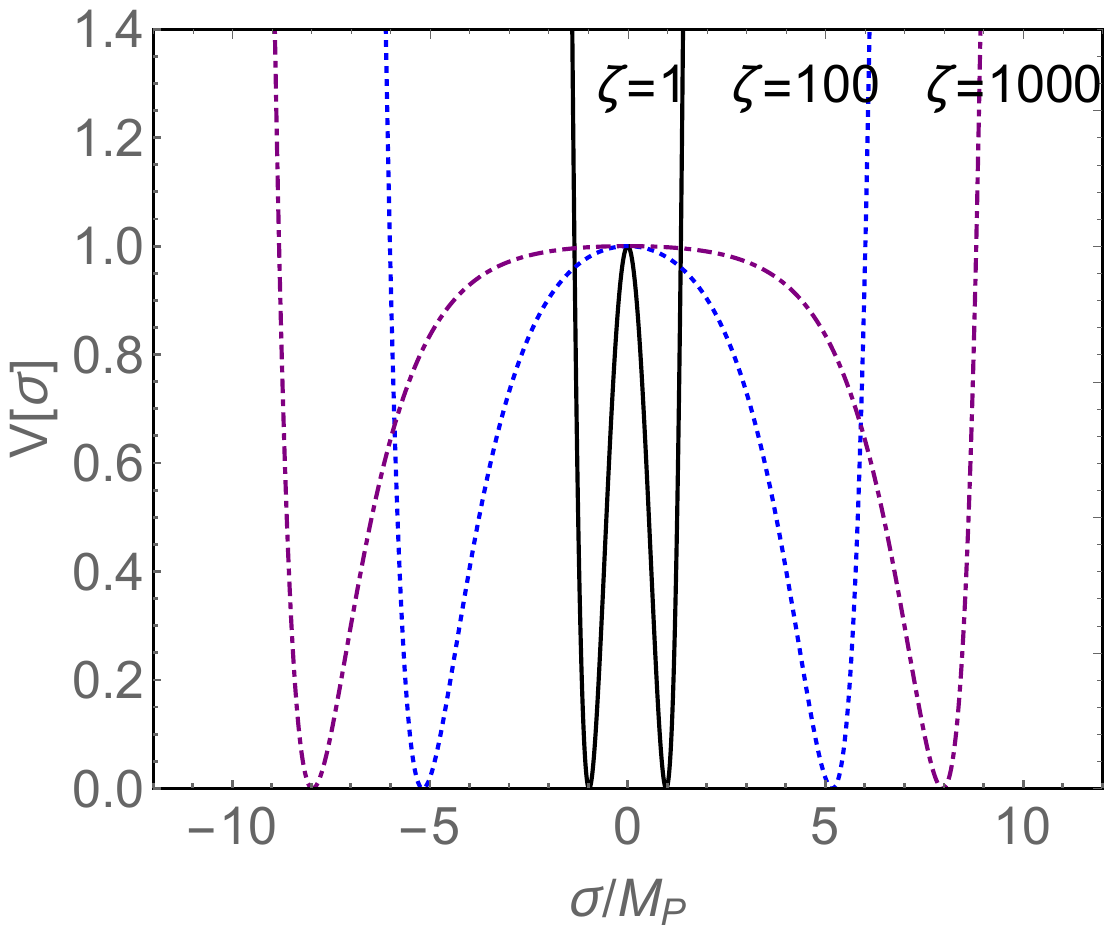}
	\includegraphics[width=0.49\textwidth,height=0.5\textwidth]{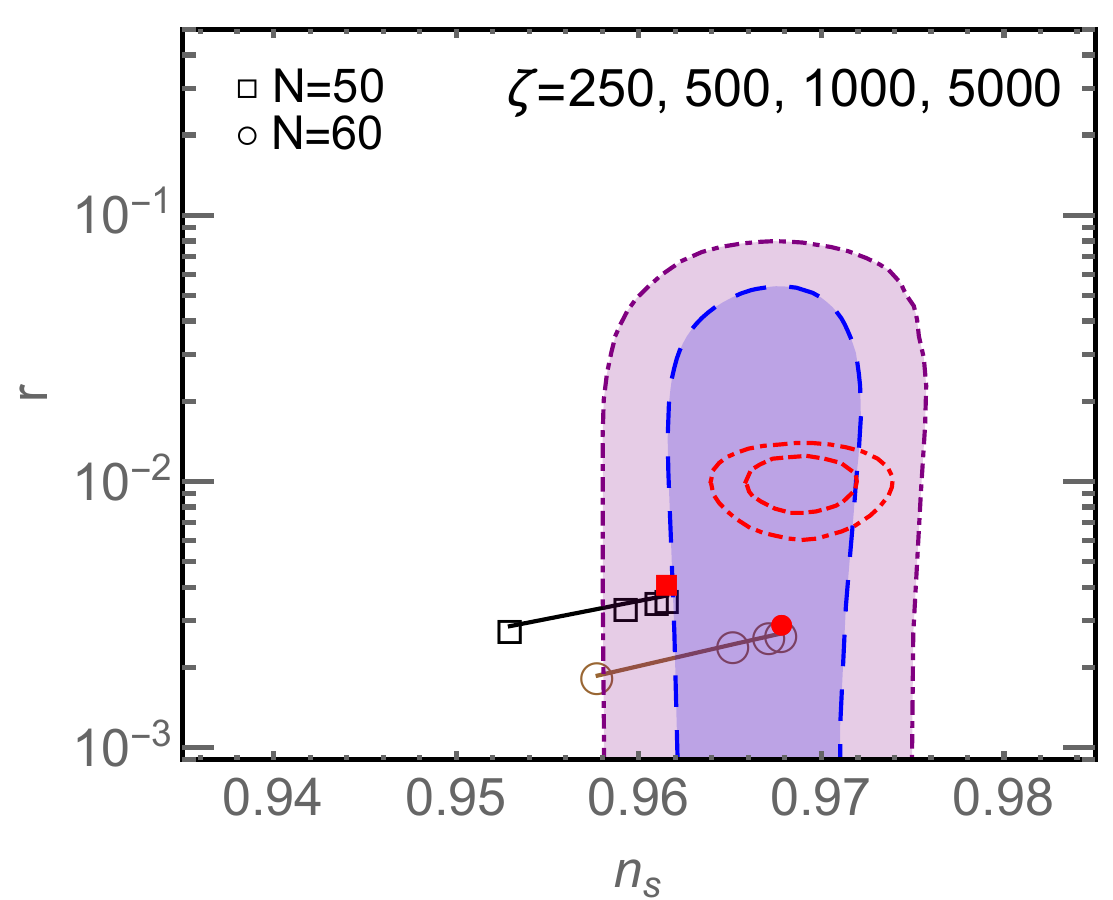}
	\caption{(Left)The scalar potentials with $\zeta=1, 100, 1000$. (Right)Values of $(n_s, r)$ for $\zeta =250, 500, 1000, 5000$ (from left to right). The predictions of $(n_s, r)$ with 4 different $\zeta$s are shown for e-folding number $N=50$ (squares) and $60$ (circles), in comparison with the shaded regions allowed by {\it Planck}~\cite{Planck:2018} with 1-$\sigma$ (blue) and 2-$\sigma$ (purple), and the future projection of CMB-S4~\cite{Abazajian:2016yjj} in red smaller contours. The filled red square and circle are the values for Starobinsky inflation.
		\label{fig:r-ns}}
\end{figure}

In the left panel of fig.~(\ref{fig:r-ns}), we show the overall shape of the potential $V(\sigma)$ with several choices of $\zeta$, $\zeta=1,100,1000$. The shape is very similar to the Higgs potential in particle physics. It is also intuitive that as $\zeta$ increases, the flat region gets broadened at small $\sigma$ in which $V(\sigma)\simeq 1$. With a flat potential, inflation may occur, starting from small $|\sigma|$ and rolling down slowly to larger $|\sigma|$ at the minimum. This picture appears different from Starobinsky model eq.~(\ref{eq:starobinsky}) where inflaton field rolls slowly from large $\sigma$ to $0$. However, as we shall show shortly, when we shift $\sigma$ with some finite value that depends on $\zeta$, the physical pictures in two models agree with each other.  

The slow-roll parameters are calculated as
\begin{align}\label{eq:slowroll}
	\epsilon &= \frac{1}{2}\left(\frac{V_\sigma }{V}\right)^2=\frac{12 \sinh ^2\left(\sqrt{\frac{2}{3}} \sigma\right)}{\left[\zeta -6 \sinh ^2\left(\frac{\sigma}{\sqrt{6}}\right)\right]^2},\\
	\eta &= \frac{V_{\sigma \sigma}}{V}=\frac{12 \cosh \left(2 \sqrt{\frac{2}{3}} \sigma\right)-4 (\zeta+3) \cosh \left(\sqrt{\frac{2}{3}} \sigma\right)}{\left[\zeta-6 \sinh ^2\left(\frac{\sigma}{\sqrt{6}}\right)\right]^2},
\end{align}
where $V_\sigma \equiv dV/d\sigma$ and $V_{\sigma\sigma}\equiv dV_\sigma/d\sigma$. 
The spectral index $n_s$ of the power spectrum and tensor-to-scalar ratio $r$ for the signal strength of primordial gravitational wave are determined by $n_s=1-6\epsilon+2\eta$ and $r=16\epsilon$. 
The theoretically calculated values of $n_s$ and $r$ for $\zeta=250,500,1000,5000$ are shown as squares ($N=50$) and circles ($N=60$) (from left to right) in the right panel of Fig.~\ref{fig:r-ns}. Here, $N$ is the e-folding number, $N\sim [50,60]$ before inflation ends,
\begin{equation}\label{eq:efold}
N\equiv \ln \frac{a_{e}}{a_{i}}\simeq \int^{t_\textrm{end}}_t Hdt \simeq \int ^{\sigma_i}_{\sigma_\textrm{e}}\frac{d\sigma}{\sqrt{2\epsilon}}, 
\end{equation}
where $a_i (a_e)$ is the scale factor at initial (end) time of the inflation, $\sigma_i (\sigma_e)$ is the corresponding field value, and $H$ is the Hubble parameter. $\sigma_e$ is determined by the violation of slow-roll condition, $\epsilon\sim 1$ or $\eta\sim 1$. 

Since there are no simple and transparent relations between $n_s$($r$) and $N$ for arbitrary $\zeta$, we numerically solve the inflationary dynamics and calculate the corresponding $n_s$ and $r$ for each $\zeta$ and $N$. As we can see in Fig.~\ref{fig:r-ns} there is viable parameter space for $\zeta$ to be consistent with the latest constraint from {\tt Planck}~\cite{Planck:2018} (color-shaded regions). The predictions of $r$ will be smaller than the sensitivity of next-generation CMB experiment~\cite{Abazajian:2016yjj} (two smaller red contours). Future upgrade with reach of $r\sim 0.003$ will be needed to detect primordial gravitational wave in this model.  

For comparison, we also show the prediction in Starobinsky inflation, $n_s\simeq 1-2/N$ and $r\simeq 12/N^2$, as the filled red square and circle. At first sight, it may seem surprising that the predictions of $n_s$ and $r$ coincide with those in Starobinsky inflation as $\zeta$ increases, after all they have different potentials and we do not know whether $6+\zeta \phi^2 \simeq \zeta \phi^2 $ is satisfied or not in the observable region, because inflation can also happen when $\phi=0$. In the following, we shall give an explanation for this coincidence through analytic analysis about the potential. First, note that the potential $V(\sigma)$ and $\phi^2(\sigma)$ are even functions of $\sigma$, so we may focus on the $\sigma < 0$ branch because the other half gives the same physical results. We can rewrite the potential as
\begin{align}\label{eq:expl}
V\left(\sigma \right)=&\frac{3}{4\alpha}\left[1-\frac{6}{\zeta}\frac{\exp\left(-\sqrt{\frac{2}{3}}\sigma \right)+\exp\left(\sqrt{\frac{2}{3}}\sigma \right)-2}{4}\right]^2\nonumber \\
= & \frac{3}{4\alpha'}\Bigg{\{} 1-\frac{3/2}{\zeta+3}\left[\exp\left(-\sqrt{\frac{2}{3}}\sigma \right)+\exp\left(\sqrt{\frac{2}{3}}\sigma \right)\right] \Bigg{\}}^2,
\end{align}
where we have defined $\alpha'=\alpha\zeta/ (\zeta +3)$ and it is evident that $\alpha'\rightarrow\alpha$ as $\zeta\rightarrow \infty$. The observable patch of the universe that went through inflation occurred when $|\sigma|$ is away from $0$. The larger $\zeta$ is, the further $|\sigma|$ is away from the origin. This may be indicated in the left plot in Fig.~(\ref{fig:r-ns}) because slow-roll conditions $\epsilon\ll 1$ and $|\eta|\ll 1$ can still be satisfied when $|\sigma|$ is far away from $0$ as $\zeta$ increases. Since we focus on $\sigma < 0$ regime for large $\zeta$, we can neglect the term $\exp\left(\sqrt{2}\sigma/\sqrt{3}\right)$ which is much smaller than $\exp\left(-\sqrt{2}\sigma/\sqrt{3}\right)$ in the relevant region. Therefore, we have the following approximation,
\begin{align}
	V\left(\sigma \right) \simeq  \frac{3}{4\alpha'} \left[ 1- \exp \left(-\sqrt{\frac{2}{3}}\left[\sigma+\sigma_0\right]\right)\right]^2,
\end{align}
where $\sigma_0$ is defined by $\exp\left(-\sqrt{\frac{2}{3}}\sigma_0\right) = \dfrac{3/2}{\zeta+3}$ in eq.~(\ref{eq:expl}). Finally making a global shift of field variable $\overline{\sigma}=\sigma+\sigma_0$, we have the same potential as the one in Starobinsky model, 
\begin{align}
	V\left(\overline{\sigma} \right) = \frac{3}{4\alpha'} \left[ 1- \exp \left(-\sqrt{\frac{2}{3}}\overline{\sigma}\right)\right]^2.
\end{align}
Now it shows that although $\sigma$ is rolling from $0$ to negative value, $\overline{\sigma}$ rolls from finite positive value to $0$, just like in Starobinksy model. The above analysis explains why the predictions of $n_s$ and $r$ are almost independent of $\zeta$ when $\zeta\gtrsim 500$, because Starobinsky model emerges as an attractor for our scenario when $\zeta$ increases. The plot also indicates that future experiments searching for primordial gravitational waves with sensitivity of tensor-to-scalar $r < 0.001$ is needed to distinguish two models.

We could also compare this model with Higgs inflation~\cite{Bezrukov:2007ep} where the standard model Higgs field acts as the inflaton thanks to its non-minimal coupling to $R$. Higgs inflation has similar dynamics as Starobinsky inflation, and same predictions for $n_s$ and $r$ at leading order if the same e-folding number is employed. This can be seen through the Lagrangian
\begin{equation}
\frac{\mathcal{L}}{\sqrt{-g}}\supset \frac{1}{2}R + \xi h^2 R - \frac{1}{2}\partial^\mu h \partial_{\mu}h - \lambda (h^2 - v^2_h)^2,
\end{equation}
where $h$ is real Higgs field and $v_h=246$GeV. In the inflationary regime $h\gg v_h$, where kinetic energy is much smaller than potential energy, one can use $v_h\simeq 0$ and neglect the kinetic term of $h$. Then solving the equation of motion of $h$, we have $h^2\simeq \xi R/(2\lambda)$. Put it back into Lagrangian, we have 	
\begin{equation}
\frac{\mathcal{L}}{\sqrt{-g}}\simeq  \frac{1}{2}R + \frac{\xi^2}{4\lambda} R^2 ,
\end{equation}
which shows that its equivalence to Starobinsky inflation.
	
One difference between Higgs inflation from our model is that $R^2$ is not included for inflation in original formalism. However, in our scenario it is necessary and crucial to include $\hat{R}^2$, without which inflation can not happen, as we have already shown in Sec.~\ref{sec:general}. This is because the breaking of local scaling symmetry without $\hat{R}^2$ would make $\phi$ as the longitudinal model of Weyl gauge boson, therefore unable to provide inflation. With $\hat{R}^2$, a new scalar degree of freedom $\chi$ appears. In the end, one scalar gets absorbed by Weyl gauge boson and the other one is responsible for inflation. 

\section{Weyl Gauge Boson as Dark Matter}\label{sec:dark}
Finally, we discuss the dual roles of Weyl gauge boson in this model. Firstly, as we have shown in previous section, the introduction of $W_\mu$ or Weyl symmetry in the formalism makes the physical theory differently from the usual $R^2$ model. The phenomenological effects on inflationary observable are also differently from the Starobinsky model, which might be tested in future precision measurements. 

Secondly, we have demonstrated that there is a $Z_2$ symmetry for the final physical massive Weyl gauge field $\overline{W}_\mu$ in the general theory of $F(\hat{R},\phi_I)$ with multiple scalars $\phi_{I}$. When $\overline{W}_\mu$ is produced in the early universe, it would be stable due to $Z_2$ symmetry and survive to present, therefore serving as a dark matter candidate. In the case that $g_W$ is extremely small, we may neglect the other interactions and only consider the gravitational production or inflation fluctuation~\cite{Graham:2015rva, Ema:2019yrd} which gives the relic abundance $\Omega_W$ for $m_W\lesssim H$, 
\begin{equation}
\Omega_W \simeq \Omega_{\textrm{DM}}\times \sqrt{\frac{m_W}{6\times 10^{-11}\GeV}}\times \left(\frac{H}{10^{13}\GeV}\right)^2,
\end{equation}
where $\Omega_{\textrm{DM}}\simeq 0.25$ and $H$ is the Hubble scale during inflation. In this case, $g_W\sim 10^{-29}$, a very tiny coupling, which indicates how challenging it is to detect such a DM particle. 

\begin{figure}[t]
	\includegraphics[width=0.65\textwidth,height=0.55\textwidth]{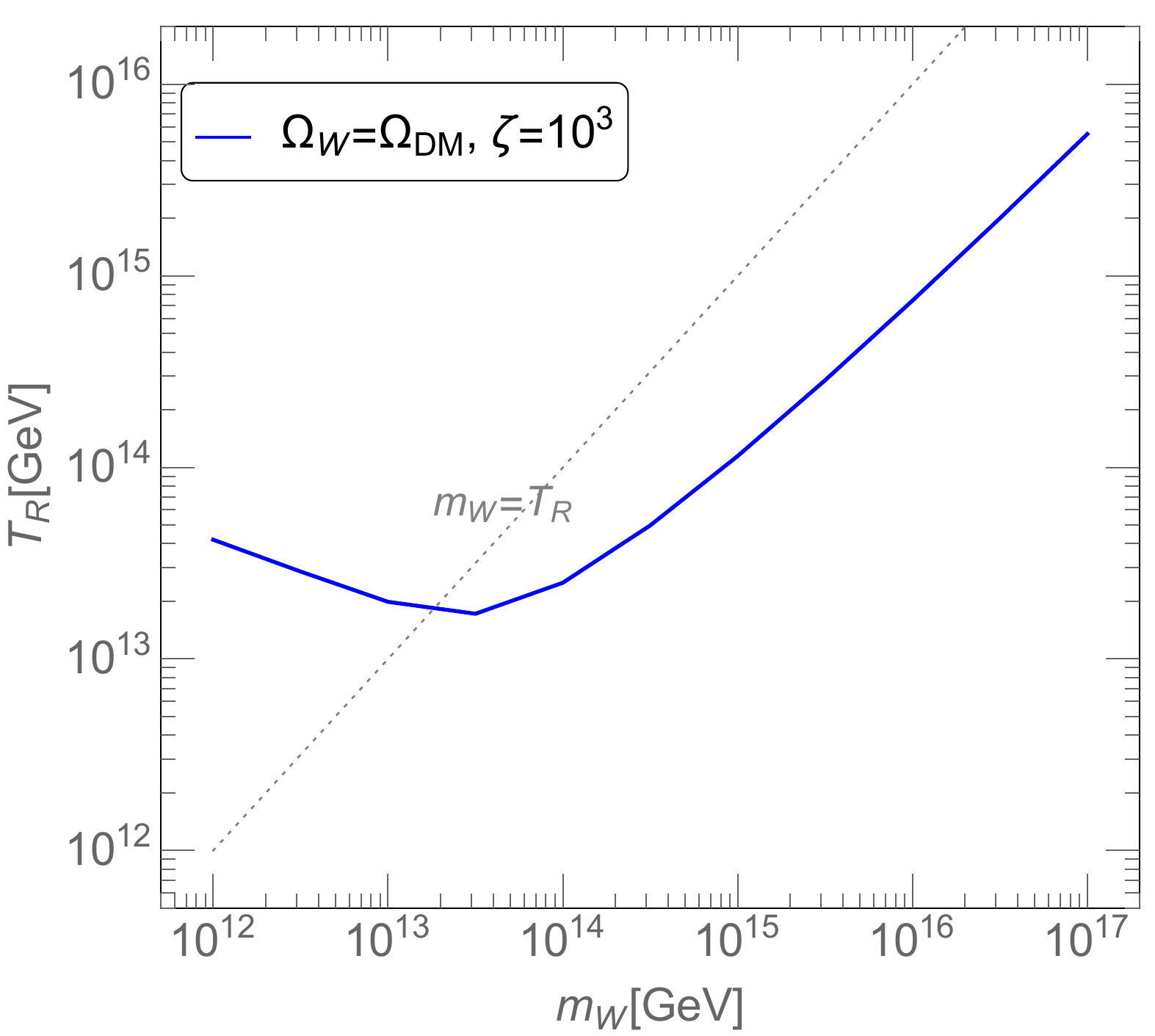}
	\caption{The solid blue curve gives the correct relic abundance $\Omega_W=\Omega_{\textrm{DM}}$ with $\zeta=10^3$, due to the scalar annihilation $\Phi + \Phi\rightarrow \overline{W}_\mu+\overline{W}_\mu$. The turnover of the blue curve at $m_W\sim T_R$ is due to suppression from Boltzmann factor, which indicates that we need to increase $T_R$ for larger $m_W$ in the $m_W>T_R$ region. 
		\label{fig:relic}}
\end{figure}

Here we provide a different mechanism to produce heavy $\overline{W}_\mu$ with $m_W \gg H$ in the early universe through the thermal production by scalar annihilations, $\Phi + \Phi\rightarrow \overline{W}_\mu+\overline{W}_\mu$, where $\Phi$ could be any scalar particles that are in the thermal equilibrium after reheating, including the standard model Higgs boson. The cross section of the annihilation process is dominated by the longitudinal mode, 
\begin{equation}
\sigma \sim \frac{g^4_W}{E^2}\frac{E^4}{m^4_W}=\frac{E^2}{(6+\zeta )^2M^4_p},
\end{equation}
where $E$ is the typical energy of final $\overline{W}$ and we have used $m_W^2=g^2_W(6+\zeta )M^2_p$ in the last equation. Interestingly, although this annihilation process is through a pure Weyl gauge interaction, the cross section is independent of gauge coupling $g_W$ but similar to that in gravitational scattering~\cite{Tang:2016vch,Tang:2017hvq}. With the cross section $\sigma$ in hand, we can calculate the yield $Y_W\equiv n_W/\bm{s}$ by integration over proper time $t$ in the early universe,  
\begin{equation}
\frac{dY_W}{d t} = \frac{T}{32\pi^4}\int ds \,\sigma\, \sqrt{s}(s-4m_\Phi^2)K_1\left(\frac{\sqrt{s}}{T}\right),
\end{equation}
where $T$ is the temperature, $n_W$ is the number density, $\bm{s}$ is the entropy density, $s$ is the square of initial energy, $m_\Phi$ is the mass of $\Phi$ and $K_1$ is the modified Bessel function of second kind with order $1$. 

We solve the above equation numerically and plot in Fig.~\ref{fig:relic} the relic density $\Omega_W=\Omega_{\textrm{DM}}$ for $\zeta=10^3$ as the blue contour in $m_W$-$T_R$ plane, where $T_R$ is the reheating temperature after inflation. The result does not depend on $m_\Phi$ as long as $m_\Phi<T_R$. Above the blue curve, we have $\Omega_W=\Omega_{\textrm{DM}}$, and $\Omega_W<\Omega_{\textrm{DM}}$ otherwise. It is seen that high $T_R$ is required to give the correct relic density. The reheating temperature $T_R$ is related to the Hubble scale $H$ and also the details of the reheating process. In the instantaneous reheating limit, we have 
\begin{equation}
	T_R\sim \sqrt{HM_p} \simeq  3\times 10^{15}\GeV\times \left(\frac{H}{10^{13}\GeV}\right)^{1/2}. 
\end{equation}
In the mean time, $H$ is determined by the energy density during inflation which is related with tensor-to-scalar ratio $r$,
\begin{equation}
H=\frac{{V}^{1/2}}{\sqrt{3}M_p}\sim 3\times 10^{13}\GeV\times \left(\frac{r}{0.01}\right)^{1/2}.
\end{equation}
Since in the considered inflation model we have $r\sim 0.003$, we are able to have $T_R\sim 4\times 10^{15}\GeV$ maximally. In such a case, $\overline{W}_\mu$ with $m_W\sim 5\times 10^{16}\GeV$ could be produced abundantly. This also indicates that the gauge coupling could be as large as $g_W\sim 10^{-3}$. 

\section{Connection to the Standard Model}
So far, we have not discussed how to embed the standard model (SM) of particle physics into the framework of scaling invariant theory. In the SM, most of the Lagrangian, including the kinetic terms of fermions and gauge fields, and Yukawa interactions, respect the scaling symmetry. The covariant derivative in the kinetic term of Higgs field $\Phi$ can be easily extended to include Weyl gauge field $W_\mu$, and the treatment is the same as our discussions in the subsection of multiple scalars. Then, there is only one dimensional parameter $\mu_\Phi$ in the Higgs potential that breaks the scaling invariance,
\begin{equation}
V(\Phi)=-\mu^2_\Phi \Phi^\dagger \Phi + \lambda_\Phi \left(\Phi^\dagger \Phi \right)^2,
\end{equation}
where $\Phi$ is the $SU(2)$ Higgs doublet. One straightforward way to promote the theory into a scaling invariant form is replacing 
\begin{equation}
\mu^2_\Phi \Phi^\dagger \Phi \rightarrow \lambda_\mu \phi^2 \Phi^\dagger \Phi,
\end{equation}
where $\phi$ is the inflation field appearing in our above discussions and $\lambda_\mu$ is a dimensionless parameter. Since $\phi^2\simeq 1$ at the potential minimum, having an electroweak-scale $\mu_\Phi$ at tree level would indicate that $\lambda_\mu$ should be very tiny, $\sim 10^{-32}$. Such tiny numbers are typical when we connect two theories with very different scales. One possible way to enhance $\lambda_\mu$ is to use the dynamical symmetry breaking and renormalization group running (RGE)~\cite{Bai:2014lea}. The basic idea is that in this framework the tree-level $\mu^2_\Phi$ is defined at Planck scale, and it can evolve to much smaller value at lower scales, see Ref.~\cite{Bai:2014lea} for details.

The above discussion is just to show that it is possible to embed SM into the framework of scaling invariance in a viable way. One may have other possible appealing mechanisms to generate the Higgs mass term, for example through the Coleman-Weinberg mechanism with a new gauge group~\cite{Iso:2009ss} and scale-invariant Lagrangian, about which a detailed discussion is beyond our scope here. 

Finally, we would like to discuss how to generate the SM particles after inflation, namely, the reheating process. From the above discussion, we know that the SM model Higgs can couple to inflaton $\phi$ or $\sigma$ field. Expanding $\sigma$ field around the minimum, we would have the cubic coupling, $\sigma \Phi^\dagger \Phi$, which induces the decay of $\sigma$ to Higgs. In such a scenario, reheating is achieved through the perturbative decay of $\sigma$. A more complicated way is to introduce an additional scalar $\varphi$ that couples to both inflaton $\phi$ and Higgs $\Phi$. Then, by adjusting the magnitude of coupling constants, the reheating temperature would be more flexible. Once Higgs particles are produced, all other SM particle can be subsequently produced through decay and annihilation. 

\section{Conclusion}\label{sec:conclusion}
We have studied inflation and dark matter in the Weyl scaling invariant $\hat{R}^2$ and general $F(\hat{R})$ gravity. The Weyl conformal symmetry is broken once the Einstein frame is chosen. We have demonstrated these theories are equivalent to Einstein gravity coupled to a scalar field and a massive Weyl gauge field. The scalar can be responsible for inflation in the early universe and the Weyl gauge boson can be a dark matter candidate.

This scenario $\hat{R}^2$ is different from the usual Starobinsky model in several ways. First, a new scalar field is introduced for preserving Weyl symmetry and the sign of its kinetic term can be both positive and negative. The resulting scalar potential is also different from that in Starobinsky model, but with the latter as an attractor or limit case in the scenario. The physical difference in inflationary observable may be tested in future experiments that aim to detect primordial gravitational waves. Secondly, the originally massless Weyl gauge boson becomes massive once the Einstein frame is fixed, after absorbing another scalar as its longitudinal model. The massive Weyl gauge boson as heavy as $5\times 10^{16}\GeV$ can be produced in the early universe through thermal annihilation, therefore serving as a dark matter candidate . 

\begin{Large}
\textbf {\newline Acknowledgments}
\end{Large}$\\$
YT is supported by National Natural Science Foundation of China (NSFC) under Grants No.~11851302 and supported by the Fundamental Research Funds for the Central Universities. YLW is supported in part by NSFC under Grants No.~11851302, No.~11851303, No.~11690022, No.~11747601, and the Strategic Priority Research Program of the Chinese Academy of Sciences under Grant No. XDB23030100 as well as the CAS Center for Excellence in Particle Physics (CCEPP).


%

\end{document}